\documentclass[journal=jctc, manuscript=article]{achemso}

\usepackage[utf8]{inputenc}
\usepackage{amsmath}
\usepackage{amsthm}
\usepackage{amssymb}
\usepackage{graphicx}
\graphicspath{{img/}}
 \DeclareGraphicsExtensions{.png,.svg}
\usepackage{braket}
\usepackage{todonotes}
\usepackage{booktabs}
\usepackage[nohyperlinks,nolist]{acronym}
\usepackage{siunitx}
\usepackage{chemformula}
\usepackage{acronym}
\usepackage{caption}
\usepackage{subcaption}
\usepackage{here}
\usepackage[version=4]{mhchem}
\usepackage[capitalize,
            noabbrev]
            {cleveref}
\makeatletter
\renewcommand{\@cite}[1]{
{$\!\! ^{(#1)}$}}
\makeatother

\newcommand{\orbital}{\ensuremath{\varphi}}

\newcommand{\nuclear}{\ensuremath{\hat{V}_{nuc}}}
\newcommand{\zorapot}{\ensuremath{\hat{V}_{Z}}}
\newcommand{\coulomb}{\ensuremath{\hat{J}}}
\newcommand{\xc}{\ensuremath{\hat{V}_{xc}}}
\newcommand{\exchange}{\ensuremath{\hat{K}}}
\newcommand{\potential}{\ensuremath{\hat{V}}}
\newcommand{\kinetic}{\ensuremath{\hat{T}}}

\newcommand{\fockMat}{\ensuremath{F}}
\newcommand{\fockOper}{\ensuremath{\hat{F}}}

\newcommand{\Helm}{\ensuremath{\hat{G}}}

\newcommand{\atomsol}{\texttt{AtomSolver}}
\newcommand{\exciting}{\texttt{exciting}}

\newcommand{\mrchem}{\texttt{MRChem}}

\title{Scalar relativistic effects with Multiwavelets: Implementation and benchmark}

\newcommand{\uit}{UiT The Arctic University of Norway, 9037 Tromsø, Norway}
\newcommand{\rice}{Department of Materials Science and NanoEngineering, Rice University, Houston, Texas 77005, United States of America.}
\newcommand{\hylleraas}{Hylleraas Centre for Quantum Molecular Sciences, \uit}
\newcommand{\chem}{Department of Chemistry, \uit}
\newcommand{\equals}{First authorship is shared between AB and SRJ}

\newcommand{\lu}{Department of Physics, University of Latvia, Jelgavas iela 3, Riga, LV-1004, Latvia}

\author{Anders Brakestad}
\affiliation{\hylleraas}
\alsoaffiliation{\chem}

\author{Stig Rune Jensen}
\affiliation{\hylleraas}
\alsoaffiliation{\chem}
\alsoaffiliation{\equals}

\author{Christian Tantardini}
\affiliation{\hylleraas}
\alsoaffiliation{\chem}
\alsoaffiliation{\rice}

\author{Quentin Pitteloud}
\affiliation{\hylleraas}
\alsoaffiliation{\chem}

\author{Peter Wind}
\affiliation{\hylleraas}
\alsoaffiliation{\chem}

\author{J{\=a}nis U\v{z}ulis}
\affiliation{\lu}

\author{Andris Gulans}
\affiliation{\lu}

\author{Kathrin Helen Hopmann}
\affiliation{\chem}

\author{Luca Frediani}
\affiliation{\hylleraas}
\alsoaffiliation{\chem}
\email{luca.frediani@uit.no}

\begin{acronym}
\acro{AO}{atomic orbital}
\acro{API}{Application Programmer Interface}
\acro{AUS}{Advanced User Support}
\acro{BO}{Born-Oppenheimer}  
\acro{CBS}{complete basis set}
\acro{CC}{Coupled Cluster}
\acro{CTCC}{Centre for Theoretical and Computational Chemistry}
\acro{CoE}{Centre of Excellence}
\acro{DC}{dielectric continuum}  
\acro{DFT}{density functional theory}  
\acro{DKH}{Douglas-Kroll-Hess}
\acro{ESC}{Elimination of Small Components}
\acro{EFP}{effective fragment potential}
\acro{ECP}{effective core potential}
\acro{EU}{European Union}
\acro{FW}{Foldy–Wouthuysen}
\acro{GGA}{generalized gradient approximation}
\acro{GPE}{Generalized Poisson Equation}
\acro{GTO}{Gaussian type orbital}
\acro{HF}{Hartree-Fock}  
\acro{HPC}{high-performance computing}
\acro{Hylleraas}[HC]{Hylleraas Centre for Quantum Molecular Sciences}
\acro{IEF}{Integral Equation Formalism}
\acro{IGLO}{individual gauge for localized orbitals}
\acro{KB}{kinetic balance}
\acro{KS}{Kohn-Sham}
\acro{LAO}{London atomic orbital}
\acro{LAPW}{linearized augmented plane wave}
\acro{LDA}{local density approximation}
\acro{MAD}{mean absolute deviation}
\acro{maxAD}{maximum absolute deviation}
\acro{MM}{molecular mechanics}  
\acro{MCSCF}{multiconfiguration self consistent field}
\acro{MPA}{multiphoton absorption}
\acro{MRA}{multiresolution analysis}
\acro{MW}{multiwavelet}
\acro{NAO}{numerical atomic orbital}
\acro{NeIC}{nordic e-infrastructure collaboration}
\acro{NMR}{nuclear magnetic resonance}
\acro{NP}{nanoparticle}  
\acro{OLED}{organic light emitting diode}
\acro{PAW}{projector augmented wave}
\acro{PBC}{Periodic Boundary Condition}
\acro{PCM}{polarizable continuum model}
\acro{PW}{plane wave}
\acro{QC}{quantum chemistry}  
\acro{QM/MM}{quantum mechanics/molecular mechanics}  
\acro{QM}{quantum mechanics}
\acro{RA}{Regular Approximation}
\acro{RCN}{Research Council of Norway}
\acro{RMSD}{root mean square deviation}
\acro{RKB}{restricted kinetic balance}
\acro{SC}{semiconductor}
\acro{SCF}{Self Consistent Field}
\acro{STSM}{short-term scientific mission}
\acro{SAPT}{symmetry-adapted perturbation theory}
\acro{SERS}{surface-enhanced raman scattering}
\acro{WPREL}[WP1]{Work Package 1}
\acro{WPROP}[WP2]{Work Package 2}
\acro{WPAPP}[WP3]{Work Package 3}
\acro{WP}{Work Package}  
\acro{X2C}{exact two-component}
\acro{ZORA}{zero-order regular approximation}
\acro{ae}{almost everywhere}
\acro{BVP}{boundary value problem}
\acro{PDE}{partial differential equation}
\acro{RDM}{1-body reduced density matrix}
\end{acronym}

\begin{document}

\maketitle  
\begin{abstract}
The importance of relativistic effects in quantum chemistry is widely recognized, not only for heavier elements but throughout the periodic table. At the same time, relativistic effect are strongest in the nuclear region, where the description of electrons through linear combination of atomic orbitals becomes more challenging. Furthermore, the choice of basis sets for heavier elements is limited compared to lighter elements where precise basis sets are available.
Thanks to the framework of multiresolution analysis, multiwavelets provide an appealing alternative to overcome this challenge: they lead to robust error control and adaptive algorithms that automatically refine the basis set description until the desired precision is reached. This allows to achieve a proper description of the nuclear region.

In this work we extended the Multiwavelet-based code \mrchem{} to the scalar \ac{ZORA} framework. We validated our implementation comparing the total energies for a small set of elements and molecules. To confirm the validity of our implementation, we compared both against a radial numerical code for atoms and the plane-wave based code \exciting{}.
\end{abstract}

\begin{tocentry}
\begin{center}
\includegraphics[width=1.0\textwidth]{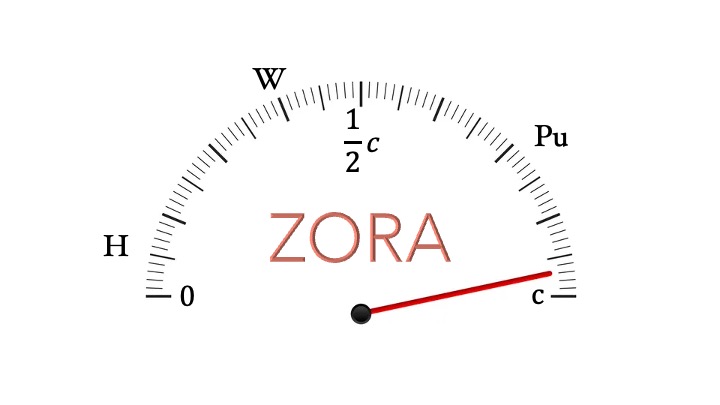}
\end{center}
\end{tocentry}

\newpage

\section{Introduction}

In his famous Nobel Price Lecture, P.A.M. Dirac stated that with the advent of quantum mechanics all fundamental problems of chemistry were in principle solved\citep{dirac1933theory}. It is however interesting that he did not seem to realize that relativity would also play a role, despite his own fundamental contribution to combining relativity and quantum mechanics\citep{Dirac_1928,10.1007/s00016-002-8369-1}. 
It is now widely accepted that the correct description of the electronic structure of atoms and molecules requires the inclusion of relativistic effects. This is particularly relevant for the core- and the innermost valence shell-electrons of heavier elements, which move at sizeable fractions of the speed of light.
A wide number of chemical properties and phenomena, such  as the yellow colour of gold\citep{Bartlett_1998,Pyykko_2012}, the liquid state of mercury\citep{Keeton_1966,Pyykko_2012}, the functioning of lead-acid batteries\citep{Ahuja_2011,Pyykko_2012}, the catalytic behavior of cobalt \citep{Pyykko_2012,Kozlowski_2016,Straka_2016,Schwerdtfeger_2010}, the difficulties to describe fluorinated compounds with DFT \citep{Boualili_2021,Mattsson_2019,Tantardini_2022} all depend on relativistic effects.
There is a wide spectrum of methods to include relativistic effects in electronic structure calculations.
They range from the use of \ac{ECP}~\citep{Hellmann_1935,Gombas_1935,Xu_2011,Stoll_2002}, also known as pseudopotentials in the solid state physics community~\citep{Heine_1970}, to the solution of the 4-component Dirac equation. In between these two extremes are a number of methods with different degrees of accuracy and complexity \citep{Saue_2011}.

The \ac{ECP} method collapses a given number of core-electrons and the nucleus to provide a core potential for the valence electron. This has the dual advantage of reducing the number of explicit electrons and at the same time including relativistic effects implicitly. On the other hand, if core electronic properties such as high pressure chemistry, core-electron spectroscopy \citep{Rolles_2018} and \ac{NMR} shielding constants \citep{Straka_2016} are considered, \ac{ECP} methods fall short, and all-electron calculations are necessary.
The four-component Dirac equation \citep{Dirac_1928} constitutes the starting point for the treatment of relativistic effects, and the full Breit Hamiltonian~\citep{Breit_1928,Breit_1932,Moss_2012,Dyall_2007,Helgaker_2012} is the the most complete treatment of relativistic effects for a many-electron system.
On the other hand, it is also the most computationally expensive: spinorbitals are 4-component complex functions.
Spin is no longer a good quantum number in a relativistic framework and the simplified picture of two electrons with opposite spin sharing the same orbital is no longer valid \citep{Jacob_2012,Saue_2011,Marian_2012}.
Most operators couple the different components of a spinorbital, leading roughly to a factor 100 in computational cost, because coupling four complex functions requires 8$\times$8 matrices instead of a real scalar \citep{Repisky_2020,Saue_2020,Zhang_2020,Storchi_2020}.
Although progress has been made to make 4-component calculations faster and easily available \citep{Repisky_2020,Saue_2020,Zhang_2020,Storchi_2020}, it is still convenient to attempt approximations that promise great reduction in the computational cost at the price of reduced accuracy.
The first step in such a hierarchy of approximations is the elimination of the small components through a \acp{FW} transformation \citep{Foldy_1950,Foldy_1952}.
The choice of transformation leads to different kinds of methods, such as the Pauli Hamiltonian \citep{Chang_1986,Kang_2004,Goldman_1977}, 
the \acp{RA} \citep{Van_1994,Dyall_1999,Mohri_2001},
the \ac{DKH} Hamiltonian \citep{Douglas_1974,Hess_1985,Hess_1986,Jansen_1989,Reiher_2006}, 
or the \ac{X2C} Hamiltonian method \citep{Kutzelnigg_2005,Liu_2006,Liu_2009,Iliavs_2007,Peng_2007,Peng_2013,Liu_2010,Saue_2011,Liu_2014,Liu_2016}. 
Further reduction to a scalar method is also possible for the Pauli Hamiltonian and the \acp{RA} \citep{Van_1994,Dyall_1999,Mohri_2001}.
In particular, the \acp{RA} has two interesting features: 
(1) the decoupling part of the \acp{FW} transformation can be expanded in a convergent series to recover the exact elimination, and
(2) the renormalization part can be exactly incorporated in the wave function. 
The \ac{ZORA}~\citep{Filatov_2003,Chang_1986,Heully_1986,Lenthe_1993,Van_1994} is the simplest form of Hamiltonian keeping only the zeroth order in both parts.
The reduction to a scalar method is carried out by applying the Dirac identity and discarding the spin-orbit term. 
The advantage of \ac{ZORA} is to keep most of the standard algorithms of quantum chemistry in their original form just re-scaling the kinetic energy by a function that includes the potential energy (see below Theory section) \citep{Filatov_2003,Chang_1986,Heully_1986,Lenthe_1993,Van_1994}.
The numerical treatment of this rescaling function can be challenging, because it displays a cusp at each nucleus\citep{Filatov_2003,Chang_1986,Heully_1986,Lenthe_1993,Van_1994}.
Standard approaches, based on atomic orbital expansions, can struggle to get an accurate description in this region.
The issue is further aggravated for heavier nuclei, where such a correction is important, and at the same time the number of basis sets available is more limited and less is known about their true precision \citep{Pantazis_2014,Guell_2008}.

Some efforts to assess the precision of \ac{ZORA} for all-electron calculations of heavier elements have been undertaken using \acp{GTO}, but it is anyway challenging to assess the precision without an external reference \citep{Pantazis_2014,Guell_2008}. 
One such option for hydrogen-like ions (\ch{He^+}, \ch{Ne^{9+}}, \ch{Ar^{17+}}, an so on) is constituted by the scaling properties of the \ac{ZORA} Hamiltonian in a two-component framework, which yields the (scaled) exact Dirac energies for such a system\citep{10.1063/1.467819}. 
However, for many-electron systems, and/or for a scalar relativistic approach, assessing the true precision of a \ac{GTO} basis is challenging and several basis sets are developed to include relativistic effects.
Examples of all-electron relativistic basis sets include the universal Gaussian basis set (UGBS) \citep{Jorge_1998}, 
the atomic natural orbitals basis sets \citep{Valera_2020,Valera_2021,Widmark_2004rano,Widmark_2004mgaa,Widmark_2005nrab}, 
the \ac{X2C} basis sets \citep{Weigend_2017}, 
and the segmented all-electron relativistic contracted (SARC) basis sets  \citep{Pantazis_2009,Pantazis_2011,Pantazis_2012,Aravena_2016,Pantazis_2020}. 
Importantly, such basis sets have to be fitted to the chosen Hamiltonian (\ac{ZORA}, \ac{DKH}).
This comes on top of the required fitting of the basis set to a given electronic structure method, and if relevant, to a particular property \citep{Pantazis_2020}. 
Although some of the known all-electron  relativistic basis sets may provide good results for a particular method and property, their transferability is limited.
The large number of available \ac{GTO} basis sets is an indication that no single basis set is good enough to describe all properties of interest to sufficient precision \citep{Guell_2008}.


In recent years, \acp{MW} \citep{Alpert_1993} have emerged as a powerful alternative to traditional local basis sets~\citep{Jensen_2017}.
Their foundation based on multiresolution analysis \citep{Keinert_2009} leads to a basis set that is not empirically parameterized.
Robust error control~\citep{Beylkin_2004, Sekino_2012, Frediani_2017} means that the user can set a finite but arbitrary target precision, and adaptive algorithms \citep{Vozovoi_2002,Perez_2008,Ruud_2013} ensure that the representation of molecular orbitals is automatically refined until the required precision is reached.
Instead of a plethora of bases to choose from, the user only needs to set the requested precision and the polynomial order of the basis, providing in practice a robust black-box, where straightforward numerical considerations guide the user's choice. \acp{MW} have proven reliable and robust in providing \ac{HF} and \ac{DFT} benchmark results for energies\cite{Jensen_2017} and properties\cite{Jensen_2016}, and have ventured both towards post-HF methods~\cite{Bischoff_2012} and relativistic treatments~\cite{Anderson_2019}.

In this work we present a \ac{MW} implementation of the \ac{ZORA} method in the \mrchem{} code.
To verify the correctness of the implementation, we consider two alternative methods: a radial atomic solver implemented as a separate package \citep{Uzulis_2022} and \ac{LAPW} implemented in the electronic-structure code \texttt{exciting}~\citep{Gulans_2014}.  
Both of these approaches provide means for systematic improvement of the precision and are capable of yielding total energies approaching the \ac{CBS} limit as demonstrated in Refs.~\citenum{Uzulis_2022} and \citenum{Gulans_2018}.

\section{Theory and implementation}
\label{sec:theory}

\subsection{From the Dirac equation to the ZORA Hamiltonian}

We will briefly expose how the \ac{ZORA} \citep{Filatov_2003,Chang_1986,Heully_1986,Lenthe_1993,Van_1994} is derived starting from the 4-component Dirac equation of an electron, which describes a 4-component spinor $\Psi$, subject to a potential $V$: 
\begin{equation}\label{eq:dirac}
    \left(
    \beta c^2 + V + c \boldsymbol{\alpha} \cdot \boldsymbol{p} 
    \right) \Psi = E \Psi,
\end{equation}
In the above equation $\boldsymbol{p}$ is the momentum operator, $c$ is the speed of light, atomic units ($m_e = 1$, $\hbar = 1$, $e = -1$) are assumed and $\boldsymbol\alpha$ and $\beta$ are defined as follows:
\begin{equation*}
    \beta = 
    \begin{pmatrix}
    I_2 & 0 \\ 0 & -I_2
    \end{pmatrix},
    \quad
    \boldsymbol{\alpha} = 
    \begin{pmatrix}
    0 & \boldsymbol{\sigma} \\ \boldsymbol{\sigma} & 0
    \end{pmatrix},
\end{equation*}

\begin{equation}
    \sigma_x = 
    \begin{pmatrix}
    0 & 1 \\ 1 & 0
    \end{pmatrix},
    \quad
    \sigma_y = 
    \begin{pmatrix}
    0 & -i \\ i & 0
    \end{pmatrix},
\quad
    \sigma_z = 
    \begin{pmatrix}
    1 & 0 \\ 0 & -1
    \end{pmatrix},
\end{equation}.

The first step towards \ac{ZORA} consists in applying the \ac{FW} transformation to the Dirac Hamiltonian of Eq.~\eqref{eq:dirac}:

\begin{equation}
    H^{FW} = U^\dagger H^{D} U,
\end{equation}
where the transformation matrix $U = W_1 W_2$ is a product of a decoupling matrix $W_1$ and a renormalization matrix $W_2$:
\begin{equation}
    W_1 = 
    \begin{pmatrix}
    1 & -R^\dagger \\  R & 1
    \end{pmatrix},
    \quad
    W_2 =
    \begin{pmatrix}
    1/\sqrt{1+R^\dagger R} & 0 \\ 0 & 1/\sqrt{1+R R^\dagger}
    \end{pmatrix},
\end{equation}
and $R$ is the exact coupling between the large and the small components of a 4-spinorbits:
\begin{equation} 
    R = \frac{1}{2c^2 - V + E} c \boldsymbol\sigma \cdot \boldsymbol{p}.
\end{equation}

The inverse potential term depends on the eigenvalue $E$, and this dependence can be expanded in a Taylor series as follows:
\begin{equation} 
    R
    = \frac{1}{2c^2 -V}\left( 1 + \frac{E}{2c^2-V} \right)^{-1} \boldsymbol\sigma \cdot \boldsymbol{p}
    = \frac{1}{2c^2 -V} \sum_{k=0}^\infty \frac{(-E)^k}{(2c^2-V)^k}\boldsymbol\sigma \cdot \boldsymbol{p}.
\end{equation}.

Restricting the expansion to the zero-th order leads to \ac{RA}.
Once the renormalization $W_2$ is also considered, the following 2-component Hamiltonian is obtained:
\begin{equation}
    H^{RA}
    =
    \frac{1}{\sqrt{1+R^\dagger R}} \left[ V + \boldsymbol\sigma \cdot \boldsymbol{p} 
    \frac{c^2}{2c^2-V} \boldsymbol\sigma \cdot \boldsymbol{p} \right] \frac{1}{\sqrt{1+R^\dagger R}}.
\end{equation}

The \ac{ZORA} Hamiltonian is finally obtained by Taylor-expanding the renormalization operator and retaining only the zero-order term:
\begin{equation}
    H^{ZORA}
    = V + \boldsymbol\sigma \cdot \boldsymbol{p} \frac{c^2}{2c^2-V} \boldsymbol\sigma \cdot \boldsymbol{p}.
\end{equation}

Using the Dirac identity
\begin{equation}
    (\boldsymbol\sigma \cdot \boldsymbol{A}) (\boldsymbol\sigma \cdot \boldsymbol{B})
    =
    \boldsymbol{A} \cdot \boldsymbol{B} + i \boldsymbol\sigma \cdot (\boldsymbol{A} \times \boldsymbol{B}),
\end{equation}
we can separate the scalar-relativistic and spin-orbit contributions to the kinetic energy covered by the first and the second terms, respectively.
Here, we keep only the scalar-relativistic part and obtain the following Hamiltonian:

\begin{equation}
    H^{ZORA}_{SR}
    = V + \boldsymbol{p} \frac{c^2}{2c^2-V} \cdot \boldsymbol{p}
    = V + \frac{1}{2} \boldsymbol{p} \kappa \cdot \boldsymbol{p} 
\end{equation}
where in the last expression we have implicitly defined $\kappa = (1-\frac{V}{2c^2})^{-1}$.
Given the \ac{ZORA} Hamiltonian, a \ac{KS} \ac{DFT} implementation is then obtained by replacing the non-relativistic kinetic energy operator with its \ac{ZORA} counterpart:
\begin{equation}
    \left(\frac{1}{2} \boldsymbol{p}\cdot \kappa \boldsymbol{p} + V \right) \orbital_i = \epsilon_i \orbital_i
\end{equation}

It should be noted that in practical implementations the potential defining $\kappa$ is usually \emph{not} the full \ac{KS} potential. Given the form of $\kappa$, the most important contribution is the nuclear attraction. Introducing Coulomb ($\coulomb$) and exchange and correlation ($\xc$) is possible, but the corresponding operator has to be recomputed numerically at each iteration. Additionally, operations such as function multiplications are difficult to perform in traditional linear combination atomic orbital basis representations.
A common choice for \ac{GTO} calculations is to use a fixed atomic potential, which is called the atomic-\ac{ZORA} approximation \citep{KNUTH201533}. The \ac{ZORA} kinetic operator can then be pre-computed and used throughout the calculation. Atomic-\ac{ZORA} has the advantage of being gauge-invariant\citep{KNUTH201533}.

\subsection{\ac{ZORA} equations in a Multiwavelet framework\label{sec:ZORA_MW} }

To obtain a \ac{MW} implementation of the \ac{ZORA} eigenvalue problem, it is necessary to transform the differential equation into an integral equation,
in analogy with the non-relativistic case \citep{Beylkin_2004,Frediani_2017}. The standard \ac{KS} equations can be concisely written as follows:
\begin{equation}\label{eq:fock_non_rel}
  \fockOper \orbital_i = \sum_j \fockMat_{ij} \orbital_j,
\end{equation}
where $\fockOper$ is the Fock operator, $\orbital_i$ refer to an occupied molecular orbital, and $\fockMat_{ij}$ are the matrix elements of the Fock operator
between two occupied orbitals, assuming a general non-canonical (non-diagonal) form. 

Within the framework of \ac{KS}-\ac{DFT}, the Fock operator consists of the kinetic energy $\kinetic$, the nuclear attraction $\nuclear$, the Coulomb repulsion
$\coulomb$, the Hartree-Fock exchange $\exchange$ scaled by some numerical factor $\lambda \in [0,1]$, and the exchange and correlation potential $\xc$:
\begin{equation}
  \fockOper = \kinetic + \nuclear + \coulomb - \lambda\exchange + \xc.
\end{equation}

In the non-relativistic domain, the coupled \ac{KS} differential equations (\ref{eq:fock_non_rel}) can be rewritten in integral form,\citep{Kalos_1962,Beylkin_2002}
by making use of the bound-state Helmholtz kernel
\begin{align}\label{eq:integral_ks_non_rel}
  \orbital_i 
  &= -2 \Helm_{\mu_i} \left( \potential\orbital_i - \sum_{j \neq i} \fockMat_{ij} \orbital_j \right),
\end{align}
where $\Helm_{\mu_i} = \big[-\nabla^2 - \mu_i^2\big]^{-1}$ is the integral convolution operator associated with the bound-state Helmholtz kernel $G(r)=e^{-{\mu_i}r}/r$, using $\mu_i = \sqrt{-2F_{ii}}$ (i.e. the diagonal elements of the Fock matrix).

In the \ac{ZORA} Hamiltonian, the kinetic energy operator becomes:
\begin{equation}
  \kinetic = \boldsymbol{p} \cdot \frac{c^2}{2c^2 - V_Z} \boldsymbol{p} = -
  \frac{c^2}{2c^2 - V_Z} \nabla^2 - \nabla \frac{c^2}{2c^2 - V_Z} \cdot
  \nabla = -\frac{1}{2} \kappa \nabla^2 - \frac{1}{2} \nabla \kappa \cdot \nabla.
\end{equation}
where $V_Z = \nuclear + \coulomb + \xc$.
Including $\coulomb$ and $\xc$ does not pose any issue in the \ac{MW} framework, other than the computational overhead of having to update the potential at every iteration, because all potentials are anyway treated on an equal footing using a numerical grid. The non-local Hartree-Fock exchange, in turn, does not seem to contribute significantly to $V_Z$ as shown in Ref.~\citenum{Faas_2000}.

Inserting the \ac{ZORA} kinetic operator into Eq.~(\ref{eq:fock_non_rel}) we obtain:
\begin{equation}\label{eq:fock_rel_1}
  \left( -\frac{1}{2} \kappa \nabla^2 - \frac{1}{2} \nabla \kappa \cdot
  \nabla + V\right) \orbital_i = \sum_j \fockMat_{ij} \orbital_j.
\end{equation}

In order to make use of the same framework as the non-relativistic implementation of Eq.~(\ref{eq:integral_ks_non_rel}), it is necessary to isolate the Laplacian and the diagonal element of the sum on the right-hand side, which together make up the bound-state Helmholtz operator $\Helm = \left(\nabla^2 + 2F_{ii} \right)^{-1} $.
This is achieved first by division by $\kappa$, recalling that $\kappa^{-1} = 1 - V_Z/2c$. The following integral equation is obtained:
\begin{equation}
\label{eq:zora_mw_integral}
  \orbital_i = -2\Helm_{\mu_i} \left[
  -\frac{1}{2}\frac{\nabla \kappa}{\kappa} \cdot \nabla \orbital_i + 
  \left(
  \frac{V}{\kappa} +
  \frac{V_Z}{2c^2} F_{ii} \right) \orbital_i -
  \frac{1}{\kappa}\sum_{j \neq i} \fockMat_{ij} \orbital_j
  \right].
\end{equation}

When $c \rightarrow \infty$, $\kappa \rightarrow 1$ and $\nabla \kappa \rightarrow 0$, and the non-relativistic form as in Eq.~(\ref{eq:integral_ks_non_rel}) is recovered. Eq.\ref{eq:zora_mw_integral} can therefore be seen as a \emph{level-shifted} version of its non-relativistic counterpart. Although it cannot be expected that the iterative solution of Eq.~(\ref{eq:zora_mw_integral}) will work for arbitrary shifts, the approach is justified by recalling that $\kappa \simeq 1$ almost everywhere, except close to the nuclei. Our tests indicate that it becomes more difficult to converge the above equation when the \ac{ZORA} contribution becomes larger, either \emph{physically} by going down the periodic table, or \emph{artificially} by letting $c \rightarrow 0$. To overcome the convergence issues for heavier nuclei (5$^{th}$ row of the periodic table) we have therefore introduced a finite nucleus model, as described in Section~\ref{sec:finite-nucleus}. For elements beyond the 5$^{th}$ row, one has to keep in mind that \ac{ZORA} becomes questionable\citep{Vissher_2000}.

An alternative approach to obtain the desired $\left(\nabla^2 + 2F_{ii}\right)$ term from Eq.~(\ref{eq:fock_rel_1}) is to add a Laplacian term ($\frac{1}{2}\nabla^2$) directly, thus avoiding division by $\kappa$. Our tests indicate that the strategy presented above works better, despite the additional singularity introduced. The main reason seems to be that the former strategy removes the Laplacian altogether, which is ill-conditioned in the discontinuous \ac{MW} basis, whereas the latter keeps part of it on the right-hand side.

\subsection{Implementation with \Aclp{MW}}

An implementation of the ZORA method as outlined above, is currently in a development version of the \mrchem{} package\citep{mrchem}, and is expected to appear in the next official release v1.2. \mrchem{} is a numerical quantum chemistry code based on a \ac{MW} framework, in which all functions and operators are represented on their own fully adaptive multi-resolution numerical real-space grid. This allows for efficient all-electron treatment of medium to large molecules (hundreds of atoms) at \ac{SCF} level of theory (both \ac{HF} and \ac{DFT}).

The $\kappa$ function is computed as a point-wise map of the chosen \ac{ZORA} potential $V_Z$ through
\begin{equation}
    \kappa(r) = \frac{1}{1 - V_Z(r)/2c^2},
\end{equation}
and similarly for its inverse
\begin{equation}
    \kappa^{-1}(r) = 1 - V_Z(r)/2c^2.
\label{eq:k-1}
\end{equation}
Both functions are represented on their own adaptive numerical grid, and subsequently treated as standard multiplicative potential operators in the SCF procedure of solving Eq.~(\ref{eq:zora_mw_integral}).

\subsection{Point nucleus models}\label{sec:point-nucleus}
In a \ac{MW} framework, the singularity of the nuclear potential can lead to numerical problems when $V_{nuc}$ is computed and used. In the non-relativistic case, the issue is circumvented by replacing the analytic $1/r$ potential with a smoothed approximation\citep{Beylkin_2004}:
\begin{equation}
    u(r) = \frac{\text{erf}(r)}{r} + \frac{1}{3\sqrt{\pi}}\left(e^{-r^2} + 16e^{-4r^2}\right),
\label{eq:point-like}
\end{equation}
and then parameterized as $u(r/s)/s$, where $s$ is a scalar smoothing parameter. The smoothing parameter depends on the nuclear charge $Z$ and the desired precision $\epsilon$ as explained in Ref~\citenum{Beylkin_2004}
\begin{equation}
    s = \left(\frac{0.00435\epsilon}{Z^5}\right)^{1/3}.
\end{equation}
The above prescription constitutes a \emph{numerical} smoothing of the nucleus to avoid accidental infinities in the representations. It should not be confused with \emph{physical} finite nucleus models \citep{Visscher_1997}, which are common in relativistic methods. This numerical smoothing is much sharper than the common finite nucleus models, and it is meant to yield results which are -- within the requested precision -- equivalent to using a point charge. 
\subsection{Finite nucleus models}\label{sec:finite-nucleus}
In order to overcome the numerical issues faced by the pointwise nuclei, we have introduced the Gaussian nuclear model, as described by Visscher~\textit{et al.}~\citep{Visscher_1997}. Not only does the model overcome the numerical problems of pointwise charges, but it is also a sounder physical description of larger nuclei.
The nuclear charge is modelled as a Gaussian distribution
\begin{equation}
    \rho^{G}(R) = \rho_{0}^{G} e^{- \xi R^{2}}
    \label{eq:gausden}
\end{equation}
with the normalisation prefactor $\rho_{0}^{G} = eZ (\xi / {\pi})^{3/2}$ and the parameter $\xi$ related to the the root-mean-square radius of the nucleus via the expression $\xi= 3 / \langle R^{2} \rangle$.
To determine $\langle R^{2} \rangle$, we apply the empirical formula for Ref.~\citenum{Johnson_1985}:
\begin{equation}
    \sqrt{\langle R^{2} \rangle} = (0.836 A^{1/3} + 0.570)fm
\end{equation}
where $A$ is the nuclear mass number and $fm$ is a femtometer length unit ($10^{-15}$~m).
We note that a choice of the nuclear mass $A$ is depends on whether the abundance of isotopes is taken in to account.
To avoid the ambiguity, we use the tabulated values of $\sqrt{\langle R^{2} \rangle}$ provided in Ref.~\citenum{Visscher_1997}.

The potential for a Gaussian charge distribution can be represented in
an analytic form by Visscher \textit{et al.}~\citep{Visscher_1997}:
\begin{equation}
\label{eq:Gauss-charge}
    V^{G}(R) = - \frac{eZ}{R} \mathrm{erf} \big ( \sqrt{\xi} R) \big),
\end{equation}
which is the solution to the Poisson equation for the charge density defined by Eq.~(\ref{eq:gausden}).

Among the different possibilities presented in Ref.~\citenum{Visscher_1997}, we have chosen the Gaussian model for the present work because it is simpler than the more realistic Fermi model and because it is implemented in all \ac{GTO} codes. The goal of the present work is the validation of the implementation. It is therefore less relevant which model is used, provided that it is consistent throughout the software packages employed.
\subsection{Methods for verification}

To verify correctness of the ZORA implementation in \mrchem{}, we use two other codes: a numerical atomic solver \citep{Uzulis_2022} and an all-electron full-potential LAPW code \exciting{}.
In this section, we briefly introduce them and provide details on the implementation of new features needed for making a direct comparison with \mrchem{}.

The atomic solver assumes atoms with spherically symmetric densities, 
and the monoelectronic wave functions are represented as $\varphi_{n\ell m}(\mathbf{r})=u_{n\ell }(r) Y_{\ell m}(\hat{r})$, where $u_{n\ell }(r)$ is a radial function defined on a one-dimensional radial grid.
The \ac{SCF} solver uses a similar approach as the \ac{MW} framework described above.
It reduces the non-relativistic Kohn-Sham equation to the integral form as follows:
\begin{equation}
\label{eq:TheOneScrPois}
    \varphi_{n\ell m}(\mathbf{r})=-2 \hat{G}_{\epsilon_{n\ell}}\left( \hat{V}\varphi_{n\ell m}(\mathbf{r})\right).
\end{equation}

This equation matches Eq.~(\ref{eq:integral_ks_non_rel}), with vanishing off-diagonal terms of the Fock matrix, because the canonical representation is used.
The radial solver originally supported only non-relativistic calculations: to implement \ac{ZORA}, we used Eq.~(\ref{eq:zora_mw_integral})
in the canonical form, i.e., replacing $F_{ii}$ with $\epsilon_{n\ell}$ and setting $F_{ij}=0$ for $i \neq j$.
This approach avoids the evaluation of second derivatives and the corresponding numerical noise which accumulates during the self-consistency iterations.

To consider systems beyond atoms, we use \texttt{exciting}.
In a nutshell, the code relies on partitioning the unit cell into non-overlapping atomic spheres and the interstitial region.
In the atomic spheres, the wavefunctions are expressed in terms of atomic-like orbitals that are updated during the self-consistence cycle.
In the interstitial region, one represents the wavefunctions with plane waves using the smoothness of the Kohn-Sham potential. 
Based on such an approach, we use two types of basis functions: (i) augmented plane-waves (APWs)\citep{Slater_1937,Andersen_1975}  and (ii) local-orbitals (LOs)\citep{Sjostedt_2000}.
Each APW combines a plane wave in the interstitial region with atomic-like orbitals in the spheres, whereas each LO is a linear combination of two atomic-like orbitals in one particular sphere and strictly zero everywhere else.
As shown in Ref.~\citenum{Gulans_2018}, it is possible to obtain a systematic convergence of the total energies simply by increasing the number of APWs and LOs. 

The \ac{ZORA} Hamiltonian is already available in the released version of the code, whereas the smeared nucleus feature described in Sec.~\ref{sec:finite-nucleus} is implemented in the present work.
Aside from adopting the electrostatic potential due to a Gaussian charge density, we also revise the evaluation of the following integral:
\begin{equation}
\label{eq:Madelung}
E_\mathrm{M}=\int \rho_\mathrm{n}(\mathbf{r}) V_\mathrm{C}(\mathbf{r}) d\mathbf{r}.
\end{equation}
If a nucleus at the site $\mathbf{R}_0$ is defined as a point charge, its density distribution is $\rho_\mathrm{n}(\mathbf{R})=Z \delta(\mathrm{R}-\mathbf{R}_0)$ leading to $E_\mathrm{M}=Z V_\mathrm{C}(\mathbf{R}_0)$.
For smeared nuclei with $\rho_\mathrm{n}(\mathbf{R})=\rho^G(\mathbf{R})$ (see Eq.~\ref{eq:Gauss-charge}), Eq.~\ref{eq:Madelung} is evaluated as an integral on a radial grid. 

\section{Computational details}
All calculations were performed with the PBE functional \citep{Perdew_1996} using the XCFun~\citep{XCFun} library, the LibXC~\citep{LibXC} library and the native implementation in \mrchem{}, the atomic solver and \exciting{}, respectively.
Being aware of slight inconsistencies in the PBE parameters employed in these libraries, we set $\beta=0.06672455060314922$ and $\mu=0.066725 \frac{\pi^2}{3}$ as defined in XCFun and use these values in calculations with all three codes.

For all \ac{MW} calculations, a development version of \mrchem{} has been employed. Interpolating polynomials of 9$^{th}$ order were used with a simulation box size of $128\,\mathrm{a_0}$. The numerical precision thresholds used are summarized in \cref{tab:mrchem_parameters}.

\begin{table}[H]
    \centering
    \caption{\mrchem{} precision parameters.}
    \label{tab:mrchem_parameters}
\begin{tabular}{l|l|l}

\toprule
Parameter & Value & Explanation \\
\midrule
\texttt{world\_prec}         & \texttt{1.0e-6}     & Overall numerical precision                    \\
\texttt{energy\_thrs}        & \texttt{1.0e-6}     & Convergence threshold total energy             \\
\texttt{orbital\_thrs}       & \texttt{1.0e-4}     & Convergence threshold maximum orbital residual \\
\bottomrule
\end{tabular}
\end{table}

All \texttt{exciting} calculations were performed using large cubic unit cell with a side length of $25\,\mathrm{a_0}$ for atoms and $30\,\mathrm{a_0}$ for molecules.
The electrostatic interaction of the periodic images was eliminated by introducing the truncation of the Coulomb interaction following the approach explained in Ref.~\citenum{Jensen_2016}.
Aside from acquiring the isolated limit, this adjustment allows us to obtain \ac{ZORA} energies consistent with both \mrchem{} and the atomic solver.
The unmodified electrostatic potential for periodic densities is defined uniquely except for an additive constant which introduces an ambiguity in the \ac{ZORA} energies~\citep{Filatov_2003,Chang_1986,Heully_1986,Lenthe_1993,Van_1994}.
The truncation of the Coulomb interaction makes the potential unique and thus removes this ambiguity completely ~\citep{Filatov_2003,Chang_1986,Heully_1986,Lenthe_1993,Van_1994}.
The canonical orbitals were expressed in terms of local orbitals and \acp{LAPW} with the cutoff $R_\mathrm{MT}K_\mathrm{max}$ sufficient to ensure a few $\mu\mathrm{Ha}$ precision.
The specific settings in the case of each atom and molecule are stored in \cref{tab:exciting_parameters} and a set of input and output data files are available in the repository. 

\begin{table}[H]
    \centering
    \caption{LAPW Parameters and structural data used in the calculations of the considered molecules and atoms. $R_{ \mathrm{MT} }^{\mathrm{min}}G_{\mathrm{max}}$ is the product of the smallest muffin-tin radius and the largest reciprocal lattice vector.
}
    \label{tab:exciting_parameters}
\begin{tabular}{c|c|c|c}

\toprule
Material & $R_{MT}^{min}G_{max}$ & $R_{MT}[a_{0}]$ & Bond length [\AA] \\
\midrule
CaO                          & 11     & 1.63 / 1.40            & 1.8221~\citep{FIPS1402}               \\
CuH                          & 10     & 1.40 / 1.00            & 1.4626~\citep{FIPS1402}               \\
SrO                          & 13     & 1.63 / 1.40            & 1.9050~\citep{Brakestad_bonds}               \\
Cu$_2$                       & 14     & 1.8                    & 2.2197~\citep{FIPS1402}               \\
AgH                          & 11     & 1.68 / 1.20            & 1.6180~\citep{FIPS1402}               \\
I$_2$                        & 14     & 1.6                    & 2.6630~\citep{Brakestad_bonds}               \\
He                           & 11     & 2                      &                      \\
Ne                           & 12     & 2                      &                      \\
Ar                           & 13     & 2                      &                      \\
Kr                           & 13     & 2                      &                      \\
Xe                           & 15     & 2                      &                      \\
\bottomrule
\end{tabular}
\end{table}

The calculations with the atomic solver employed a 5$^{th}$ order polynomial on a radial grid with the innermost and outermost points $r_\mathrm{min}=10^{-8} \,\mathrm{a_0}$ and $r_\mathrm{max}=35\,\mathrm{a_0}$.
The number of radial points was set to 5000, which is fully sufficient to guarantee a sub-$\mu\mathrm{Ha}$ precision. 

The total energies of diatomic molecules were calculated in \mrchem{} and \exciting{} without geometry relaxation using the internuclear distances given in Table~\ref{tab:exciting_parameters}.
The bond lengths in Table~\ref{tab:exciting_parameters}, are taken from the NIST~\citep{FIPS1402} database with the exception of SrO and I$_2$, which are private communication from not published work.

\section{Results and discussion}

\subsection{Relative contributions of the \ac{ZORA} terms}

The current \mrchem{} implementation is able to include all local contributions of the potential into \zorapot{}. It is therefore possible to measure their relative weights for a given atom and how their contribution changes with the nuclear charge. We have performed a series of calculations for the noble gases from helium to xenon, with all seven possibilities: one contribution only, two contributions and all three. 
The results are summarized in Fig.~\ref{fig:relative_contributions}. The nuclear potential \nuclear{} is the largest contribution, as expected. It is followed by the Coulomb term \coulomb{}, and the exchange and correlation potential \xc{} is the smallest one.
Moreover the relativistic correction increases roughly with the fourth power of the nuclear charge as expected\cite{Pyykko_2012}, and the nuclear term becomes progressively more dominant for heavier atoms.

\newcommand{\WIDTH}{0.32}
\begin{figure}[H]
    \centering
        \begin{subfigure}[b]{\WIDTH\textwidth}
            \centering
            \includegraphics[width=\textwidth]{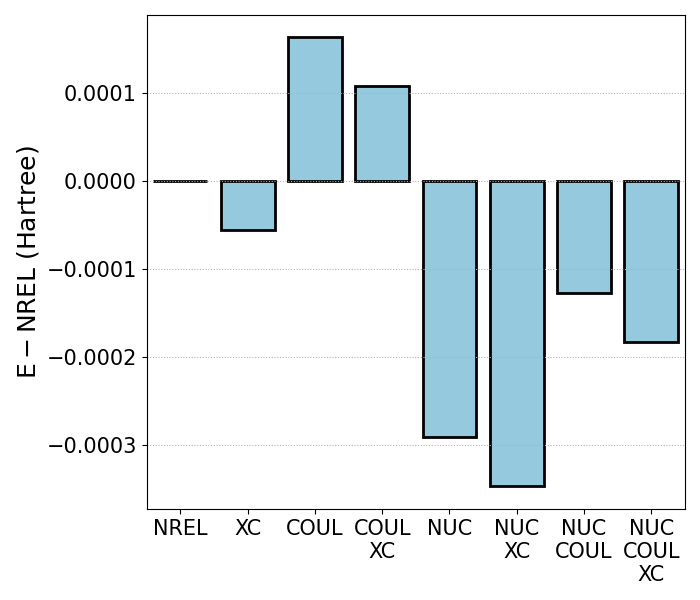}
            \caption{Helium}
            \label{subfig:relcont_He}
        \end{subfigure}
    \hfill
        \begin{subfigure}[b]{\WIDTH\textwidth}
            \centering
            \includegraphics[width=\textwidth]{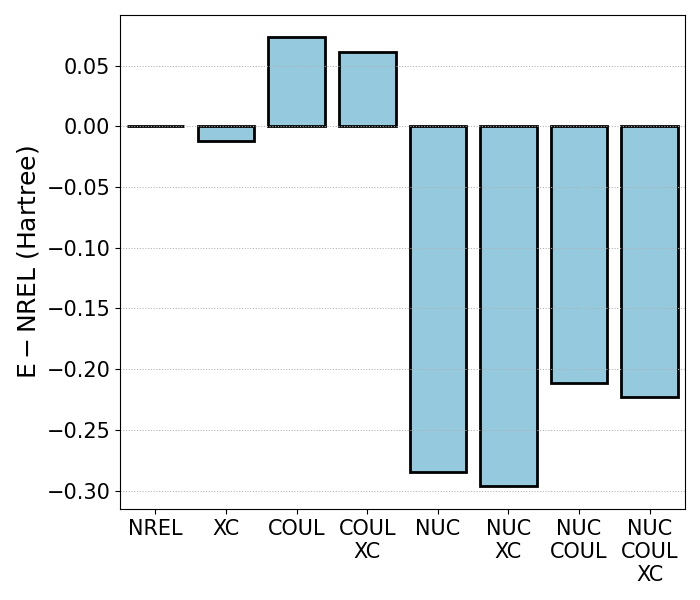}
            \caption{Neon}
            \label{subfig:relcont_Ne}
        \end{subfigure}
    \hfill
        \begin{subfigure}[b]{\WIDTH\textwidth}
            \centering
            \includegraphics[width=\textwidth]{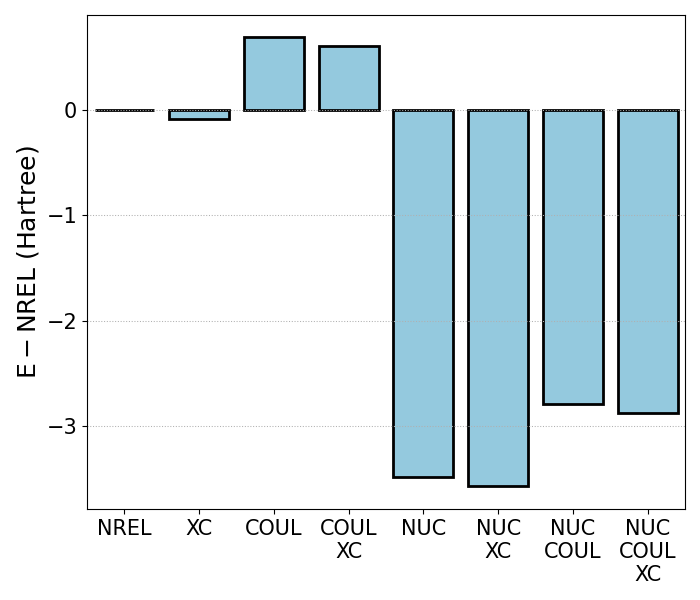}
            \caption{Argon}
            \label{subfig:relcont_Ar}
        \end{subfigure}
    \\
        \begin{subfigure}[b]{\WIDTH\textwidth}
            \centering
            \includegraphics[width=\textwidth]{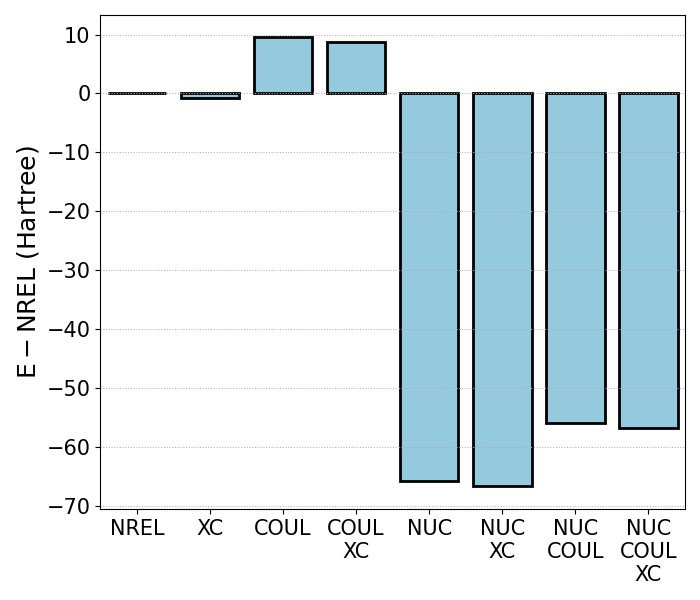}
            \caption{Krypton}
            \label{subfig:relcont_Kr}
        \end{subfigure}
        \begin{subfigure}[b]{\WIDTH\textwidth}
            \centering
            \includegraphics[width=\textwidth]{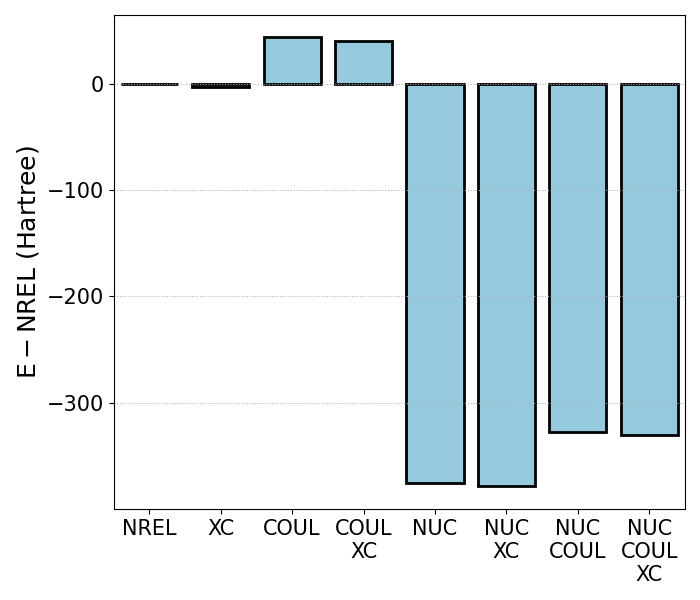}
            \caption{Xenon}
            \label{subfig:relcont_Xe}
        \end{subfigure}
    \caption{Relative contributions to the total energy (in Hartree) for all seven possible \ac{ZORA} operators, compared to the non-relativistic total energy, for the noble gases He -- Xe. Note that the scales on the y-axes are different for each sub-plot.}
    \label{fig:relative_contributions}
\end{figure}

\subsection{Validation}
To validate the \ac{ZORA} implementation in \mrchem{}, we perform total energy calculations of noble gas atoms and a small set of molecules broadly covering the first five rows of the periodic table (H-Xe).
In the case of the atoms, we apply three different types of calculations: the atomic solver, \acp{LAPW} and \acp{MW}.
For the diatomic molecules we compare the results obtained using \ac{MW} and \ac{LAPW} by means of \mrchem{} and \exciting{}, respectively.

To assess the potential agreement which can be achieved between the various codes, we have performed non-relativistic calculations with all three codes, using point-charge nuclei (numerically smoothed as described in Section \ref{sec:point-nucleus} in the case of \acp{MW}). The results are summarized in~\cref{tab:validation-nr}.
We have then obtained a \ac{RMSD} of the relative error between \acp{MW} and \acp{LAPW} equal to $5.21\cdot10^{-8}$, and between \acp{MW} and atomic solver equal to $8.07\cdot10^{-10}$.
We concluded that the three methods are in very good agreement, setting the mark for what can be expected in the \ac{ZORA} domain, using a Gaussian nuclear charge distribution as in~\cref{eq:Gauss-charge}. 
The \ac{ZORA} results are summarized in~\cref{tab:validation-ZORA} showing a \ac{RMSD} for the relative errors between \acp{MW} and \acp{LAPW} equal to $3.95\cdot 10^{-8}$, and between \acp{MW} and atomic solver equal to $6.93\cdot10^{-9}$, in line with what has been observed for the non-relativistic regime, thus confirming the validity of the implementation.
In terms of absolute errors, we find that most discrepancies in the total energies are within 10~microHartrees, and only in two cases (AgH and I$_2$) the differences are larger, yet they do not exceed 21~microHartrees.
This level of agreement is well below the so called \emph{chemical accuracy threshold} (1~kcal/mol). 

Finally, we find that introducing the relativistic corrections in the multiwavelet formalism as expressed in Eq.~\ref{eq:zora_mw_integral} leads to a minor increase in \mrchem{} runtimes.
Therefore we anticipate that the analysis of the performance (runtimes and parallel scaling) given in Ref.~\cite{Wind2023} remains valid in the ZORA case.

\begin{table}[H]
    \centering
    \caption{Non-relativistic total energies (given in Hartrees) obtained using three different codes and their relative differences. 
    In all cases, the point-like nucleus model is used.}
    \label{tab:validation-nr}
\begin{tabular}{l | rrrrr}
\toprule
        &            &             &           & \multicolumn{2}{c}{ Relative difference}  \\
Species & \atomsol{} & \exciting{} & \mrchem{} & \mrchem{} vs & \mrchem{} vs \\
        &            &             &           & \atomsol{}  & \exciting{} \\
\midrule
He     &      -2.892935499   &      -2.892935485   &      -2.892935497 &  -4.1e-10   & 4.4e-09 \\
Ne     &    -128.866433504   &    -128.866433222   &    -128.866433465 &  -3.1e-10   & 1.9e-09 \\
Ar     &    -527.346141146   &    -527.346138312   &    -527.346140237 &  -1.7e-09   & 3.7e-09 \\
Kr     &   -2753.416138121   &   -2753.416137130   &   -2753.416137698 &  -0.1e-09   & 0.2e-09 \\
Xe     &   -7234.233259457   &   -7234.233258510   &   -7234.233259210 &  -0.03e-09   & 0.09e-09 \\
\midrule
CaO    &        -         &   -752.562050270   &     -752.562058080   &       -     & 1.0e-08 \\
CuH    &        -         &  -1640.901716750   &    -1640.901726830   &       -     & 6.1e-09 \\
SrO    &        -         &  -3208.161711950   &    -3208.161721780   &       -     & 3.1e-09 \\
Cu$_2$ &        -         &  -3280.675839820   &    -3280.675845520   &       -     & 1.7e-09 \\
AgH    &        -         &  -5200.162245050   &    -5200.162259390   &       -     & 2.8e-09 \\
I$_2$  &        -         & -13840.158307100   &   -13840.158328190   &       -     & 1.5e-09 \\
\bottomrule
\end{tabular}
\end{table}

\begin{table}[H]
    \centering
    \caption{Scalar-relativistic ZORA total energies (given in Hartrees) obtained using three different codes and their relative differences. In all cases, the smeared nucleus model is used.}
    \label{tab:validation-ZORA}
\begin{tabular}{l | rrrrr}
\toprule
        &            &             &           & \multicolumn{2}{c}{ Relative difference}  \\
Species & \atomsol{} & \exciting{} & \mrchem{} & \mrchem{} vs & \mrchem{} vs \\
        &            &             &           & \atomsol{}  & \exciting{} \\
\midrule
He     &     -2.893118631     &     -2.893118615  &      -2.89311862 &  -2.2e-09   &  3.2e-09  \\
Ne     &   -129.089335224     &   -129.089335102  &    -129.08933496 &  -2.1e-09   & -1.1e-09  \\
Ar     &   -530.224039154     &   -530.224036217  &    -530.22403777 &  -2.6e-09   &  2.9e-09  \\
Kr     &  -2810.049763806     &  -2810.049761830  &   -2810.04975504 &  -3.1e-09   & -2.4e-09  \\
Xe     &  -7564.596665119     &  -7564.596661910  &   -7564.59655428 &  -1.5e-08   & -1.4e-08  \\
\midrule
CaO    &       -          &    -757.195492387  &    -757.195491088 &       -     & -1.7e-09  \\
CuH    &       -          &   -1663.238492560  &   -1663.238503660 &       -     &  6.7e-09  \\
SrO    &       -          &   -3279.826587400  &   -3279.826596402 &       -     &  2.7e-09  \\
Cu$_2$ &       -          &   -3325.342851280  &   -3325.342843306 &       -     & -2.4e-09  \\
AgH    &       -          &   -5380.042000350  &   -5380.042003292 &       -     &  0.5e-09  \\
I$_2$  &       -          &  -14448.747786900  &  -14448.747761310 &       -     & -1.8e-09  \\
\bottomrule
\end{tabular}
\end{table}

\section{Conclusions}
We have formulated the \ac{ZORA} method in a form compatible with multiwavelets and implemented it in the \mrchem{} program. The validity and precision of the implementation has been tested against a radial, numerical atomic code and a plane wave code, showing excellent agreement.
The current study was done with the specific idea to validate method and theory with a small benchmark: atoms and diatomics, covering broadly the periodic table up to and including $5^{th}$-row elements.
The model is also capable of dealing with all parts of the electronic potential self-consistently, with the exception of the \ac{HF} exchange, which is the only one which cannot be expressed in closed form.

\begin{acknowledgement}
This work was supported by the Norwegian Research Council through a Centre of Excellence grant (Hylleraas Centre 262695), a FRIPRO grant (ReMRChem 324590), by the Tromsø Research Foundation (TFS2016KHH), and by UNINETT Sigma2 through grants of computer time (nn9330k and nn4654k). J{\=a}nis U\v{z}ulis acknowledges funding provided by the project "Strengthening of the capacity of doctoral studies at the University of Latvia within the framework of the new doctoral model”, identification No. 8.2.2.0/20/I/006.
Andris Gulans acknowledges funding provided by European Regional Development Fund via the Central Finance and Contracting Agency of Republic of Latvia under the grant agreement 1.1.1.5/21/A/004.
We would like to thank Dr.~Susi Lehtola from Department of Chemistry at the University of Helsinki in Finland for the useful feedback to the original draft of the manuscript.
\end{acknowledgement}

\bibliography{references} 

\end{document}